\documentclass[12pt,a4paper]{article}

\makeatletter
\newcount\@hour\newcount\@minute\chardef\@x10\chardef\@xv60
\def\tcitime{
\def\@time{%
  \@minute\time\@hour\@minute\divide\@hour\@xv
  \ifnum\@hour<\@x 0\fi\the\@hour:%
  \multiply\@hour\@xv\advance\@minute-\@hour
  \ifnum\@minute<\@x 0\fi\the\@minute
  }}%
\def\QCTOpt[#1]#2{%
  \def\QCTOptB{#1}
  \def\QCTOptA{#2}
}
\def\QCTNOpt#1{%
  \def\QCTOptA{#1}
  \let\QCTOptB\empty
}
\def\Qct{%
  \@ifnextchar[{%
    \QCTOpt}{\QCTNOpt}
}
\def\QCBOpt[#1]#2{%
  \def\QCBOptB{#1}
  \def\QCBOptA{#2}
}
\def\QCBNOpt#1{%
  \def\QCBOptA{#1}
  \let\QCBOptB\empty
}
\def\Qcb{%
  \@ifnextchar[{%
    \QCBOpt}{\QCBNOpt}
}
\def\PrepCapArgs{%
  \ifx\QCBOptA\empty
    \ifx\QCTOptA\empty
      {}%
    \else
      \ifx\QCTOptB\empty
        {\QCTOptA}%
      \else
        [\QCTOptB]{\QCTOptA}%
      \fi
    \fi
  \else
    \ifx\QCBOptA\empty
      {}%
    \else
      \ifx\QCBOptB\empty
        {\QCBOptA}%
      \else
        [\QCBOptB]{\QCBOptA}%
      \fi
    \fi
  \fi
}
\newcount\GRAPHICSTYPE
\GRAPHICSTYPE=\z@
\def\GRAPHICSPS#1{%
 \ifcase\GRAPHICSTYPE
   \special{ps: #1}%
 \or
   \special{language "PS", include "#1"}%
 \fi
}%
\def\graffile#1#2#3#4{%
    \leavevmode
    \raise -#4 \BOXTHEFRAME{%
        \hbox to #2{\raise #3\hbox{\null #1}}}%
}%
\def\draftbox#1#2#3#4{%
 \leavevmode\raise -#4 \hbox{%
  \frame{\rlap{\protect\tiny #1}\hbox to #2%
   {\vrule height#3 width\z@ depth\z@\hfil}%
  }%
 }%
}%
\newcount\draft
\draft=\z@

\newif\ifwasdraft
\wasdraftfalse

\def\GRAPHIC#1#2#3#4#5{%
 \ifnum\draft=\@ne\draftbox{#2}{#3}{#4}{#5}%
  \else\graffile{#1}{#3}{#4}{#5}%
  \fi
 }%
\def\addtoLaTeXparams#1{%
    \edef\LaTeXparams{\LaTeXparams #1}}%
\newif\ifBoxFrame \BoxFramefalse
\newif\ifOverFrame \OverFramefalse
\newif\ifUnderFrame \UnderFramefalse

\def\BOXTHEFRAME#1{%
   \hbox{%
      \ifBoxFrame
         \frame{#1}%
      \else
         {#1}%
      \fi
   }%
}

\def\doFRAMEparams#1{\BoxFramefalse\OverFramefalse\UnderFramefalse\readFRAMEparams#1\end}%
\def\readFRAMEparams#1{%
 \ifx#1\end%
  \let\next=\relax
  \else
  \ifx#1i\dispkind=\z@\fi
  \ifx#1d\dispkind=\@ne\fi
  \ifx#1f\dispkind=\tw@\fi
  \ifx#1t\addtoLaTeXparams{t}\fi
  \ifx#1b\addtoLaTeXparams{b}\fi
  \ifx#1p\addtoLaTeXparams{p}\fi
  \ifx#1h\addtoLaTeXparams{h}\fi
  \ifx#1X\BoxFrametrue\fi
  \ifx#1O\OverFrametrue\fi
  \ifx#1U\UnderFrametrue\fi
  \ifx#1w
    \ifnum\draft=1\wasdrafttrue\else\wasdraftfalse\fi
    \draft=\@ne
  \fi
  \let\next=\readFRAMEparams
  \fi
 \next
 }%
\def\IFRAME#1#2#3#4#5#6{%
      \bgroup
      \let\QCTOptA\empty
      \let\QCTOptB\empty
      \let\QCBOptA\empty
      \let\QCBOptB\empty
      #6%
      \parindent=0pt%
      \leftskip=0pt
      \rightskip=0pt
      \setbox0 = \hbox{\QCBOptA}%
      \@tempdima = #1\relax
      \ifOverFrame
          \typeout{This is not implemented yet}%
          \show\HELP
      \else
         \ifdim\wd0>\@tempdima
            \advance\@tempdima by \@tempdima
            \ifdim\wd0 >\@tempdima
               \textwidth=\@tempdima
               \setbox1 =\vbox{%
                  \noindent\hbox to \@tempdima{\hfill\GRAPHIC{#5}{#4}{#1}{#2}{#3}\hfill}\\%
                  \noindent\hbox to \@tempdima{\parbox[b]{\@tempdima}{\QCBOptA}}%
               }%
               \wd1=\@tempdima
            \else
               \textwidth=\wd0
               \setbox1 =\vbox{%
                 \noindent\hbox to \wd0{\hfill\GRAPHIC{#5}{#4}{#1}{#2}{#3}\hfill}\\%
                 \noindent\hbox{\QCBOptA}%
               }%
               \wd1=\wd0
            \fi
         \else
            \ifdim\wd0>0pt
              \hsize=\@tempdima
              \setbox1 =\vbox{%
                \unskip\GRAPHIC{#5}{#4}{#1}{#2}{0pt}%
                \break
                \unskip\hbox to \@tempdima{\hfill \QCBOptA\hfill}%
              }%
              \wd1=\@tempdima
           \else
              \hsize=\@tempdima
              \setbox1 =\vbox{%
                \unskip\GRAPHIC{#5}{#4}{#1}{#2}{0pt}%
              }%
              \wd1=\@tempdima
           \fi
         \fi
         \@tempdimb=\ht1
         \advance\@tempdimb by \dp1
         \advance\@tempdimb by -#2%
         \advance\@tempdimb by #3%
         \leavevmode
         \raise -\@tempdimb \hbox{\box1}%
      \fi
      \egroup%
}%
\def\DFRAME#1#2#3#4#5{%
 \begin{center}
     \let\QCTOptA\empty
     \let\QCTOptB\empty
     \let\QCBOptA\empty
     \let\QCBOptB\empty
     \ifOverFrame 
        #5\QCTOptA\par
     \fi
     \GRAPHIC{#4}{#3}{#1}{#2}{\z@}
     \ifUnderFrame 
        \par #5\QCBOptA
     \fi
 \end{center}%
 }%
\def\FFRAME#1#2#3#4#5#6#7{%
 \begin{figure}[#1]%
  \let\QCTOptA\empty
  \let\QCTOptB\empty
  \let\QCBOptA\empty
  \let\QCBOptB\empty
  \ifOverFrame
    #4
    \ifx\QCTOptA\empty
    \else
      \ifx\QCTOptB\empty
        \caption{\QCTOptA}%
      \else
        \caption[\QCTOptB]{\QCTOptA}%
      \fi
    \fi
    \ifUnderFrame\else
      \label{#5}%
    \fi
  \else
    \UnderFrametrue%
  \fi
  \begin{center}\GRAPHIC{#7}{#6}{#2}{#3}{\z@}\end{center}%
  \ifUnderFrame
    #4
    \ifx\QCBOptA\empty
      \caption{}%
    \else
      \ifx\QCBOptB\empty
        \caption{\QCBOptA}%
      \else
        \caption[\QCBOptB]{\QCBOptA}%
      \fi
    \fi
    \label{#5}%
  \fi
  \end{figure}%
 }%
\newcount\dispkind%
\def\FRAME#1#2#3#4#5#6#7#8{%
 \ifnum\draft=\@ne
   \wasdrafttrue
 \else
   \wasdraftfalse%
 \fi
 \def\LaTeXparams{}%
 \dispkind=\z@
 \def\LaTeXparams{}%
 \doFRAMEparams{#1}%
 \ifnum\dispkind=\z@\IFRAME{#2}{#3}{#4}{#7}{#8}{#5}\else
  \ifnum\dispkind=\@ne\DFRAME{#2}{#3}{#7}{#8}{#5}\else
   \ifnum\dispkind=\tw@
    \edef\@tempa{\noexpand\FFRAME{\LaTeXparams}}%
    \@tempa{#2}{#3}{#5}{#6}{#7}{#8}%
    \fi
   \fi
  \fi
  \ifwasdraft\draft=1\else\draft=0\fi{}%
 }%
\def\TEXUX#1{"texux"}

\def\limfunc#1{\mathop{\rm #1}}%

\long\def\QQQ#1#2{%
     \long\expandafter\def\csname#1\endcsname{#2}}%
\@ifundefined{QTP}{\def\QTP#1{}}{}
\@ifundefined{Qlb}{}{}
\@ifundefined{Qlt}{}{}
\long\def\QQA#1#2{}%
\def\QTR#1#2{{\csname#1\endcsname #2}}
\def\EXPAND#1[#2]#3{}%
\def\NOEXPAND#1[#2]#3{}%
\def\LaTeXparent#1{}%
\def\ChildStyles#1{}%
\def\ChildDefaults#1{}%
\def\QTagDef#1#2#3{}%
\def\QQfnmark#1{\footnotemark}

\@ifundefined{INDEX}{\def\INDEX#1#2{}{}}{}%
\@ifundefined{SUBINDEX}{\def\SUBINDEX#1#2#3{}{}{}}{}%
\def\initial#1{\bigbreak{\raggedright\large\bf #1}\kern 2\p@
   \penalty3000}%
\@ifundefined{ZZZ}{}{\makeatletter\input gnuindex.sty\makeatother\makeindex\makeatletter}%
\@ifundefined{abstract}{%
 \def\abstract{%
  \if@twocolumn
   \section*{Abstract (Not appropriate in this style!)}%
   \else \small 
   \begin{center}{\bf Abstract\vspace{-.5em}\vspace{\z@}}\end{center}%
   \quotation 
   \fi
  }%
 }{%
 }%
\@ifundefined{endabstract}{\def\endabstract
  {\if@twocolumn\else\endquotation\fi}}{}%
\@ifundefined{maketitle}{\def\maketitle#1{}}{}%
\@ifundefined{affiliation}{\def\affiliation#1{}}{}%
\@ifundefined{proof}{}{}%
\@ifundefined{endproof}{}{}%
\@ifundefined{newfield}{\def\newfield#1#2{}}{}%
\@ifundefined{chapter}{\def\chapter#1{\par(Chapter head:)#1\par }%
 \newcount\c@chapter}{}%
\@ifundefined{part}{\def\part#1{\par(Part head:)#1\par }}{}%
\@ifundefined{section}{\def\section#1{\par(Section head:)#1\par }}{}%
\@ifundefined{subsection}{\def\subsection#1%
 {\par(Subsection head:)#1\par }}{}%
\@ifundefined{subsubsection}{\def\subsubsection#1%
 {\par(Subsubsection head:)#1\par }}{}%
\@ifundefined{paragraph}{\def\paragraph#1%
 {\par(Subsubsubsection head:)#1\par }}{}%
\@ifundefined{subparagraph}{\def\subparagraph#1%
 {\par(Subsubsubsubsection head:)#1\par }}{}%
\@ifundefined{therefore}{}{}%
\@ifundefined{backepsilon}{}{}%
\@ifundefined{yen}{}{}%
\@ifundefined{registered}{%
   \def\registered{\relax\ifmmode{}\r@gistered
                    \else$\m@th\r@gistered$\fi}%
 \def\r@gistered{^{\ooalign
  {\hfil\raise.07ex\hbox{$\scriptstyle\rm\text{R}$}\hfil\crcr
  \mathhexbox20D}}}}{}%
\@ifundefined{Eth}{}{}%
\@ifundefined{eth}{}{}%
\@ifundefined{Thorn}{}{}%
\@ifundefined{thorn}{}{}%
\@ifundefined{degree}{}{}%
\newdimen\theight
\def\Column{%
 \vadjust{\setbox\z@=\hbox{\scriptsize\quad\quad tcol}%
  \theight=\ht\z@\advance\theight by \dp\z@\advance\theight by \lineskip
  \kern -\theight \vbox to \theight{%
   \rightline{\rlap{\box\z@}}%
   \vss
   }%
  }%
 }%
\def\qed{%
 \ifhmode\unskip\nobreak\fi\ifmmode\ifinner\else\hskip5\p@\fi\fi
 \hbox{\hskip5\p@\vrule width4\p@ height6\p@ depth1.5\p@\hskip\p@}%
 }%
\def\miss{\hbox{\vrule height2\p@ width 2\p@ depth\z@}}%
\def\tcol#1{{\baselineskip=6\p@ \vcenter{#1}} \Column}  %
\def\newfmtname{LaTeX2e}
\def\chkcompat{%
   \if@compatibility
   \else
     \usepackage{latexsym}
   \fi
}

\ifx\fmtname\newfmtname
  \DeclareOldFontCommand{\rm}{\normalfont\rmfamily}{\mathrm}
  \DeclareOldFontCommand{\sf}{\normalfont\sffamily}{\mathsf}
  \DeclareOldFontCommand{\tt}{\normalfont\ttfamily}{\mathtt}
  \DeclareOldFontCommand{\bf}{\normalfont\bfseries}{\mathbf}
  \DeclareOldFontCommand{\it}{\normalfont\itshape}{\mathit}
  \DeclareOldFontCommand{\sl}{\normalfont\slshape}{\@nomath\sl}
  \DeclareOldFontCommand{\sc}{\normalfont\scshape}{\@nomath\sc}
  \chkcompat
\fi

\@ifundefined{theorem}{}{}
\@ifundefined{lemma}{}{}
\@ifundefined{corollary}{}{}
\@ifundefined{conjecture}{}{}
\@ifundefined{proposition}{}{}
\@ifundefined{axiom}{}{}
\@ifundefined{remark}{}{}
\@ifundefined{example}{}{}
\@ifundefined{exercise}{}{}
\@ifundefined{definition}{}{}

\@ifundefined{mathletters}{%
  \newcounter{equationnumber}  
  \def\mathletters{%
     \addtocounter{equation}{1}
     \edef\@currentlabel{\theequation}%
     \setcounter{equationnumber}{\c@equation}
     \setcounter{equation}{0}%
     \edef\theequation{\@currentlabel\noexpand\alph{equation}}%
  }
  
}{}

\@ifundefined{BibTeX}{%
    \def\BibTeX{{\rm B\kern-.05em{\sc i\kern-.025em b}\kern-.08em
                 T\kern-.1667em\lower.7ex\hbox{E}\kern-.125emX}}}{}%
\@ifundefined{AmS}%
    {\def\AmS{{\protect\usefont{OMS}{cmsy}{m}{n}%
                A\kern-.1667em\lower.5ex\hbox{M}\kern-.125emS}}}{}%
\@ifundefined{AmSTeX}{}{}%
\let\DOTSI\relax
\def\RIfM@{\relax\ifmmode}%
\def\FN@{\futurelet\next}%
\newcount\intno@
\def\iint{\DOTSI\intno@\tw@\FN@\ints@}%
\def\iiint{\DOTSI\intno@\thr@@\FN@\ints@}%
\def\iiiint{\DOTSI\intno@4 \FN@\ints@}%
\def\idotsint{\DOTSI\intno@\z@\FN@\ints@}%
\def\ints@{\findlimits@\ints@@}%
\newif\iflimtoken@
\newif\iflimits@
\def\findlimits@{\limtoken@true\ifx\next\limits\limits@true
 \else\ifx\next\nolimits\limits@false\else
 \limtoken@false\ifx\ilimits@\nolimits\limits@false\else
 \ifinner\limits@false\else\limits@true\fi\fi\fi\fi}%
\def\multint@{\int\ifnum\intno@=\z@\intdots@                          
 \else\intkern@\fi                                                    
 \ifnum\intno@>\tw@\int\intkern@\fi                                   
 \ifnum\intno@>\thr@@\int\intkern@\fi                                 
 \int}
\def\multintlimits@{\intop\ifnum\intno@=\z@\intdots@\else\intkern@\fi
 \ifnum\intno@>\tw@\intop\intkern@\fi
 \ifnum\intno@>\thr@@\intop\intkern@\fi\intop}%
\def\intic@{%
    \mathchoice{\hskip.5em}{\hskip.4em}{\hskip.4em}{\hskip.4em}}%
\def\negintic@{\mathchoice
 {\hskip-.5em}{\hskip-.4em}{\hskip-.4em}{\hskip-.4em}}%
\def\ints@@{\iflimtoken@                                              
 \def\ints@@@{\iflimits@\negintic@
   \mathop{\intic@\multintlimits@}\limits                             
  \else\multint@\nolimits\fi                                          
  \eat@}
 \else                                                                
 \def\ints@@@{\iflimits@\negintic@
  \mathop{\intic@\multintlimits@}\limits\else
  \multint@\nolimits\fi}\fi\ints@@@}%
\def\intkern@{\mathchoice{\!\!\!}{\!\!}{\!\!}{\!\!}}%
\def\plaincdots@{\mathinner{\cdotp\cdotp\cdotp}}%
\def\intdots@{\mathchoice{\plaincdots@}%
 {{\cdotp}\mkern1.5mu{\cdotp}\mkern1.5mu{\cdotp}}%
 {{\cdotp}\mkern1mu{\cdotp}\mkern1mu{\cdotp}}%
 {{\cdotp}\mkern1mu{\cdotp}\mkern1mu{\cdotp}}}%
\def\RIfM@{\relax\protect\ifmmode}
\def\text{\RIfM@\expandafter\text@\else\expandafter\mbox\fi}
\let\nfss@text\text
\def\text@#1{\mathchoice
   {\textdef@\displaystyle\f@size{#1}}%
   {\textdef@\textstyle\tf@size{\firstchoice@false #1}}%
   {\textdef@\textstyle\sf@size{\firstchoice@false #1}}%
   {\textdef@\textstyle \ssf@size{\firstchoice@false #1}}%
   \glb@settings}

\def\textdef@#1#2#3{\hbox{{%
                    \everymath{#1}%
                    \let\f@size#2\selectfont
                    #3}}}
\newif\iffirstchoice@
\firstchoice@true
\def\Let@{\relax\iffalse{\fi\let\\=\cr\iffalse}\fi}%
\def\vspace@{\def\vspace##1{\crcr\noalign{\vskip##1\relax}}}%
\def\multilimits@{\bgroup\vspace@\Let@
 \baselineskip\fontdimen10 \scriptfont\tw@
 \advance\baselineskip\fontdimen12 \scriptfont\tw@
 \lineskip\thr@@\fontdimen8 \scriptfont\thr@@
 \lineskiplimit\lineskip
 \vbox\bgroup\ialign\bgroup\hfil$\m@th\scriptstyle{##}$\hfil\crcr}%
\def\Sb{_\multilimits@}%
\def\endSb{\crcr\egroup\egroup\egroup}%
\def\Sp{^\multilimits@}%

\newdimen\ex@
\def\rightarrowfill@#1{$#1\m@th\mathord-\mkern-6mu\cleaders
 \hbox{$#1\mkern-2mu\mathord-\mkern-2mu$}\hfill
 \mkern-6mu\mathord\rightarrow$}%
\def\leftarrowfill@#1{$#1\m@th\mathord\leftarrow\mkern-6mu\cleaders
 \hbox{$#1\mkern-2mu\mathord-\mkern-2mu$}\hfill\mkern-6mu\mathord-$}%
\def\leftrightarrowfill@#1{$#1\m@th\mathord\leftarrow
\mkern-6mu\cleaders
 \hbox{$#1\mkern-2mu\mathord-\mkern-2mu$}\hfill
 \mkern-6mu\mathord\rightarrow$}%
\def\overrightarrow{\mathpalette\overrightarrow@}%
\def\overrightarrow@#1#2{\vbox{\ialign{##\crcr\rightarrowfill@#1\crcr
 \noalign{\kern-\ex@\nointerlineskip}$\m@th\hfil#1#2\hfil$\crcr}}}%

\def\overleftarrow{\mathpalette\overleftarrow@}%
\def\overleftarrow@#1#2{\vbox{\ialign{##\crcr\leftarrowfill@#1\crcr
 \noalign{\kern-\ex@\nointerlineskip}$\m@th\hfil#1#2\hfil$\crcr}}}%
\def\overleftrightarrow{\mathpalette\overleftrightarrow@}%
\def\overleftrightarrow@#1#2{\vbox{\ialign{##\crcr
   \leftrightarrowfill@#1\crcr
 \noalign{\kern-\ex@\nointerlineskip}$\m@th\hfil#1#2\hfil$\crcr}}}%
\def\underrightarrow{\mathpalette\underrightarrow@}%
\def\underrightarrow@#1#2{\vtop{\ialign{##\crcr$\m@th\hfil#1#2\hfil
  $\crcr\noalign{\nointerlineskip}\rightarrowfill@#1\crcr}}}%

\def\underleftarrow{\mathpalette\underleftarrow@}%
\def\underleftarrow@#1#2{\vtop{\ialign{##\crcr$\m@th\hfil#1#2\hfil
  $\crcr\noalign{\nointerlineskip}\leftarrowfill@#1\crcr}}}%
\def\underleftrightarrow{\mathpalette\underleftrightarrow@}%
\def\underleftrightarrow@#1#2{\vtop{\ialign{##\crcr$\m@th
  \hfil#1#2\hfil$\crcr
 \noalign{\nointerlineskip}\leftrightarrowfill@#1\crcr}}}%

\def\qopnamewl@#1{\mathop{\operator@font#1}\nlimits@}
\let\nlimits@\displaylimits
\def\setboxz@h{\setbox\z@\hbox}

\def\varlim@#1#2{\mathop{\vtop{\ialign{##\crcr
 \hfil$#1\m@th\operator@font lim$\hfil\crcr
 \noalign{\nointerlineskip}#2#1\crcr
 \noalign{\nointerlineskip\kern-\ex@}\crcr}}}}

 \def\rightarrowfill@#1{\m@th\setboxz@h{$#1-$}\ht\z@\z@
  $#1\copy\z@\mkern-6mu\cleaders
  \hbox{$#1\mkern-2mu\box\z@\mkern-2mu$}\hfill
  \mkern-6mu\mathord\rightarrow$}
\def\leftarrowfill@#1{\m@th\setboxz@h{$#1-$}\ht\z@\z@
  $#1\mathord\leftarrow\mkern-6mu\cleaders
  \hbox{$#1\mkern-2mu\copy\z@\mkern-2mu$}\hfill
  \mkern-6mu\box\z@$}

\def\projlim{\qopnamewl@{proj\,lim}}
\def\injlim{\qopnamewl@{inj\,lim}}
\def\varinjlim{\mathpalette\varlim@\rightarrowfill@}
\def\varprojlim{\mathpalette\varlim@\leftarrowfill@}
\def\varliminf{\mathpalette\varliminf@{}}
\def\varliminf@#1{\mathop{\underline{\vrule\@depth.2\ex@\@width\z@
   \hbox{$#1\m@th\operator@font lim$}}}}
\def\varlimsup{\mathpalette\varlimsup@{}}
\def\varlimsup@#1{\mathop{\overline
  {\hbox{$#1\m@th\operator@font lim$}}}}

\def\stackunder#1#2{\mathrel{\mathop{#2}\limits_{#1}}}%
\begingroup \catcode `|=0 \catcode `[= 1
\catcode`]=2 \catcode `\{=12 \catcode `\}=12
\catcode`\\=12 
|gdef|@alignverbatim#1\end{align}[#1|end[align]]
|gdef|@salignverbatim#1\end{align*}[#1|end[align*]]

|gdef|@alignatverbatim#1\end{alignat}[#1|end[alignat]]
|gdef|@salignatverbatim#1\end{alignat*}[#1|end[alignat*]]

|gdef|@xalignatverbatim#1\end{xalignat}[#1|end[xalignat]]
|gdef|@sxalignatverbatim#1\end{xalignat*}[#1|end[xalignat*]]

|gdef|@gatherverbatim#1\end{gather}[#1|end[gather]]
|gdef|@sgatherverbatim#1\end{gather*}[#1|end[gather*]]

|gdef|@gatherverbatim#1\end{gather}[#1|end[gather]]
|gdef|@sgatherverbatim#1\end{gather*}[#1|end[gather*]]

|gdef|@multilineverbatim#1\end{multiline}[#1|end[multiline]]
|gdef|@smultilineverbatim#1\end{multiline*}[#1|end[multiline*]]

|gdef|@arraxverbatim#1\end{arrax}[#1|end[arrax]]
|gdef|@sarraxverbatim#1\end{arrax*}[#1|end[arrax*]]

|gdef|@tabulaxverbatim#1\end{tabulax}[#1|end[tabulax]]
|gdef|@stabulaxverbatim#1\end{tabulax*}[#1|end[tabulax*]]

|endgroup

\def\align{\@verbatim \frenchspacing\@vobeyspaces \@alignverbatim
You are using the "align" environment in a style in which it is not defined.}

\@namedef{align*}{\@verbatim\@salignverbatim
You are using the "align*" environment in a style in which it is not defined.}
\expandafter\let\csname endalign*\endcsname =\endtrivlist

\def\alignat{\@verbatim \frenchspacing\@vobeyspaces \@alignatverbatim
You are using the "alignat" environment in a style in which it is not defined.}

\@namedef{alignat*}{\@verbatim\@salignatverbatim
You are using the "alignat*" environment in a style in which it is not defined.}
\expandafter\let\csname endalignat*\endcsname =\endtrivlist

\def\xalignat{\@verbatim \frenchspacing\@vobeyspaces \@xalignatverbatim
You are using the "xalignat" environment in a style in which it is not defined.}

\@namedef{xalignat*}{\@verbatim\@sxalignatverbatim
You are using the "xalignat*" environment in a style in which it is not defined.}
\expandafter\let\csname endxalignat*\endcsname =\endtrivlist

\def\gather{\@verbatim \frenchspacing\@vobeyspaces \@gatherverbatim
You are using the "gather" environment in a style in which it is not defined.}

\@namedef{gather*}{\@verbatim\@sgatherverbatim
You are using the "gather*" environment in a style in which it is not defined.}
\expandafter\let\csname endgather*\endcsname =\endtrivlist

\def\multiline{\@verbatim \frenchspacing\@vobeyspaces \@multilineverbatim
You are using the "multiline" environment in a style in which it is not defined.}

\@namedef{multiline*}{\@verbatim\@smultilineverbatim
You are using the "multiline*" environment in a style in which it is not defined.}
\expandafter\let\csname endmultiline*\endcsname =\endtrivlist

\def\arrax{\@verbatim \frenchspacing\@vobeyspaces \@arraxverbatim
You are using a type of "array" construct that is only allowed in AmS-LaTeX.}

\def\tabulax{\@verbatim \frenchspacing\@vobeyspaces \@tabulaxverbatim
You are using a type of "tabular" construct that is only allowed in AmS-LaTeX.}

\@namedef{arrax*}{\@verbatim\@sarraxverbatim
You are using a type of "array*" construct that is only allowed in AmS-LaTeX.}
\expandafter\let\csname endarrax*\endcsname =\endtrivlist

\@namedef{tabulax*}{\@verbatim\@stabulaxverbatim
You are using a type of "tabular*" construct that is only allowed in AmS-LaTeX.}
\expandafter\let\csname endtabulax*\endcsname =\endtrivlist

\def\@@eqncr{\let\@tempa\relax
    \ifcase\@eqcnt \def\@tempa{& & &}\or \def\@tempa{& &}%
      \else \def\@tempa{&}\fi
     \@tempa
     \if@eqnsw
        \iftag@
           \@taggnum
        \else
           \@eqnnum\stepcounter{equation}%
        \fi
     \fi
     \global\tag@false
     \global\@eqnswtrue
     \global\@eqcnt\z@\cr}

 \def\endequation{%
     \ifmmode\ifinner 
      \iftag@
        \addtocounter{equation}{-1} 
        $\hfil
           \displaywidth\linewidth\@taggnum\egroup \endtrivlist
        \global\tag@false
        \global\@ignoretrue   
      \else
        $\hfil
           \displaywidth\linewidth\@eqnnum\egroup \endtrivlist
        \global\tag@false
        \global\@ignoretrue 
      \fi
     \else   
      \iftag@
        \addtocounter{equation}{-1} 
        \eqno \hbox{\@taggnum}
        \global\tag@false%
        $$\global\@ignoretrue
      \else
        \eqno \hbox{\@eqnnum}
        $$\global\@ignoretrue
      \fi
     \fi\fi
 } 

 \newif\iftag@ \tag@false
 
 \def\tag{\@ifnextchar*{\@tagstar}{\@tag}}
 \def\@tag#1{%
     \global\tag@true
     \global\def\@taggnum{(#1)}}
 \def\@tagstar*#1{%
     \global\tag@true
     \global\def\@taggnum{#1}%
}

\makeatother

\begin{document}

\begin{center}
\vspace{1pt}\textbf{DESCRIPTION OF THE LOW-ENERGY DOUBLET NEUTRON-DEUTERON
SCATTERING ON THE BASIS OF THE TRITON BOUND AND VIRTUAL STATE PARAMETERS}

\textbf{\ }

\vspace{1pt}V. A. Babenko \footnote{{\normalsize
\mbox{E-mail address:
pet@online.com.ua}}} and N. M. Petrov

\textit{Bogolyubov Institute for Theoretical Physics,}

\textit{National Academy of Sciences of Ukraine, Kiev, Ukraine}
\end{center}

\noindent Low-energy doublet neutron-deuteron scattering is described on the
basis of the triton bound and virtual state parameters --- the energies and
the nuclear vertex constants of these states. The van Oers-Seagrave formula
is derived from the Bargmann representation of the $S$ matrix for a system
having two states. The presence of a pole in this formula is shown to be a
direct corollary of the existence of a low-energy triton virtual state.
Simple explicit expressions for the $nd$ scattering length and for the pole
of the function $k\cot \delta $ are obtained in terms of the triton bound
and virtual state parameters. Numerical calculations of the $nd$ low-energy
scattering parameters show their high sensitivity to variations in the
asymptotic normalization constant of the virtual state $C_{v}^{2}$\thinspace
. The $C_{v}^{2}$ value fitted in our model to the experimental result for
the $nd$ scattering length is $C_{v}^{2}=0.0592$.\vspace{1pt}

\noindent

\begin{center}
\textbf{1. INTRODUCTION}
\end{center}

\vspace{1pt}

The low-energy characteristics of three-body systems were studied previously
\lbrack 1, 2\rbrack\ on the basis of the two-body model with the Hulth\'{e}n
potential; in particular, a correlation between the binding energy of three
hadrons and the hadron-deuteron scattering length was analyzed over a wide
region of three-body parameters. An important role of the virtual triton
state was revealed in studying this dependence (Phillips line) in the region
of the experimental values of the triton binding energy $E_{T}$ and the
doublet $nd$ scattering length $^{2}a_{nd}$. The characteristics associated
with the $T\rightarrow d+n$ decay vertex were also calculated for the bound (%
$T$) and virtual ($v$) triton states. The results of the calculations for
the position of the virtual triton level, $B_{v}$, as reckoned from the
threshold of the elastic $nd$ channel and for nuclear vertex constants $%
G_{T}^{2}$ and $G_{v}^{2}$ characterizing the ground and the virtual triton
state, respectively, agree with the relevant results of three-body
calculations and with experimental data. The results from \lbrack 1,
2\rbrack\ evince a clear-cut correlation between the low-energy
characteristics of two- and three-hadron systems.

\vspace{1pt}

It was established in \lbrack 3--6\rbrack\ that the effective-range
expansion of the doublet $S$-wave phase shift for $nd$ scattering involves a
pole situated in the nonphysical region near the threshold. For the function
$k\cot \delta $, the corresponding four-parameter representation involving a
pole is referred to as the van Oers-Seagrave formula. It should be noted
that this formula was found in a purely empirical way without any
theoretical justification.

\vspace{1pt}

In order to analyze two- and three-hadron systems, we propose here an
approach based on the $S$ matrix corresponding to so-called Bargmann
potentials \lbrack 7, 8\rbrack . We will consider the relation between the $%
S $-matrix representation introduced by Bargmann \lbrack 7, 8\rbrack\ and
the effective-range expansion and demonstrate that, on the basis of this
representation, the effective-range approximation can be obtained for the
case where the system in question has one state, bound or virtual. An
example of such a situation is provided by neutron-proton scattering. We
will then generalize this analysis to systems having two states, as in the
case of doublet neutron-deuteron scattering. It will be shown that the van
Oers-Seagrave formula for the function $k\cot \delta $ involving a pole
follows directly from the Bargmann representation of the $S$ matrix for a
system having two states. This analysis makes it possible to relate the
parameters of the effective-range expansion to the characteristics of the
bound and virtual states of the system.

\vspace{1pt}

We assume that any state of the system is specified by two parameters, the
energy corresponding to the pole of the $S$ matrix on the imaginary axis in
the complex plane of the wave number $k$ and the nuclear vertex constant,
which is directly expressed in terms of the residue of the $S$ matrix at
this pole \lbrack 9\rbrack . Nuclear vertex constants are fundamental
physical characteristics of nuclei like more conventional quantities,
including mass, spin, and parity. Kinematical factors apart, nuclear vertex
constants are related to the on-shell amplitude for the virtual or real
decay (or fusion) of a nucleus into two fragments. The properties of vertex
constants, their values for a number of nuclei, and methods for determining
them experimentally and theoretically were surveyed elsewhere \lbrack
9\rbrack .

\newpage

\begin{center}
\textbf{2. DERIVATION OF THE EFFECTIVE-RANGE APPROXIMATION FROM THE BARGMANN
REPRESENTATION OF THE S MATRIX INVOLVING ONE STATE: NEUTRON-PROTON SYSTEM}
\end{center}

\vspace{1pt}

Bargmann \lbrack 7\rbrack\ proposed taking the Jost function $f\left(
k\right) $ entering into the well-known $S$-matrix expression \lbrack
8\rbrack
\begin{equation}
S\left( k\right) =\frac{f\left( -k\right) }{f\left( k\right) }  \tag{1}
\end{equation}
in the form of a rational function having some simple poles and zeros and
exhibiting a correct asymptotic behavior at high energies --- that is,
approaching unity at infinity:
\begin{equation}
\lim_{k\rightarrow \infty }f\left( k\right) =1\,.  \tag{2}
\end{equation}
In the simplest case of only one state in the system, the Jost function has
the form
\begin{equation}
f\left( k\right) =\frac{k-i\alpha }{k+i\lambda }\,,  \tag{3}
\end{equation}
where the parameter $\lambda $ is always positive, whereas the parameter $%
\alpha $ is positive for a bound state and negative for a virtual state.
Upon the substitution of (3) into (1), the $S$ matrix for the system having
one state assumes the form
\begin{equation}
S\left( k\right) =\frac{k+i\alpha }{k-i\alpha }\,\frac{k+i\lambda }{%
k-i\lambda }\,.  \tag{4}
\end{equation}
This state has the energy
\begin{equation}
E_{0}=-\frac{\hbar ^{2}\alpha ^{2}}{2m}\,,  \tag{5}
\end{equation}
where $m$ is the reduced mass of the system and $\hbar $ is the Planck
constant. The first pole factor in expression (4) for the $S$ matrix
corresponds to a physical bound or a virtual state of the system. The second
factor in the $S$ matrix (4) includes the well-known redundant pole \lbrack
8, 10\rbrack , which is associated with no bound state of the system. The
redundant pole ensures the correct asymptotic behavior (2) of the Jost
function. The nuclear vertex constant $G^{2}$ for the state being considered
is directly expressed in terms of the residue of the $S$ matrix at the pole $%
k=i\alpha $ as
\begin{equation}
G^{2}=i\pi \rule[5pt]{4.6pt}{0.5pt}\hspace{-0.5em}\lambda ^{2}\stackunder{%
k=i\alpha }{\limfunc{Res}}S\left( k\right) \,,  \tag{6}
\end{equation}
where $\rule[5pt]{4.6pt}{0.5pt}\hspace{-0.5em}\lambda \equiv \hbar /mc$ is
the reduced Compton wavelength of the system. Calculating the residue on the
basis of (6) and (4), we express the vertex constant in terms of the
parameters $\alpha $ and $\lambda $ as
\begin{equation}
G^{2}=-2\pi \rule[5pt]{4.6pt}{0.5pt}\hspace{-0.5em}\lambda ^{2}\alpha \frac{%
\alpha +\lambda }{\alpha -\lambda }\,.  \tag{7}
\end{equation}

\vspace{1pt}

Using (4) and considering that the $S$ matrix is expressed in terms of the
phase shift $\delta \left( k\right) $ as
\begin{equation}
S\left( k\right) =e^{2i\delta \left( k\right) }=\frac{\cot \delta +i}{\cot
\delta -i}\,,  \tag{8}
\end{equation}
we represent the function $k\cot \delta $ in the form
\begin{equation}
k\cot \delta =-\frac{1}{a}+\frac{1}{2}r_{e}k^{2}\,.  \tag{9}
\end{equation}
This is the expression of the effective-range approximation, with the
scattering length $a$ and the effective range $r_{e}$ being given by
\begin{equation}
a=\frac{1}{\alpha }+\frac{1}{\lambda }\,,  \tag{10}
\end{equation}
\begin{equation}
r_{e}=\frac{2}{\alpha +\lambda }\,.  \tag{11}
\end{equation}
Thus, the effective-range approximation (9) immediately follows from the
Bargmann representation (4) of the $S$ matrix for a system having one state.
Formulas (10) and (11) relate the parameters of the effective-range
approximation to the $S$-matrix parameters and, together with (5) and (7),
the low-energy scattering parameters to the binding energy $E_{0}$ and the
nuclear vertex constant $G^{2}$ (parameters of the bound state of the
system).

\vspace{1pt}

As a specific example, we will now consider neutron-proton scattering in the
triplet spin state. In this case, the system features one bound state, the
deuteron, with the binding energy being $\varepsilon _{d}=2.225\,MeV$. The
nuclear vertex constant $G_{d}^{2}$ for the deuteron corresponds to the $%
d\rightarrow n+p$ vertex and takes the value of $G_{d}^{2}=0.43\,fm$ \lbrack
9\rbrack . It should be recalled that the nuclear vertex constants are
directly related to the asymptotic normalization factors for the bound-state
wave functions. The latter factors are often introduced in the analysis
along with the nuclear vertex constants. In the case under consideration,
the constant $G_{d}^{2}$ is expressed in terms of the dimensionless
asymptotic normalization factor $C_{d}$ for the deuteron wave function as
\begin{equation}
G_{d}^{2}=2\pi \rule[5pt]{4.6pt}{0.5pt}\hspace{-0.5em}\lambda ^{2}\alpha
C_{d}^{2}\,,  \tag{12}
\end{equation}
where $\rule[5pt]{4.6pt}{0.5pt}\hspace{-0.5em}\lambda =2%
\rule[5pt]{4.6pt}{0.5pt}\hspace{-0.5em}\lambda _{N}\,$, with $%
\rule[5pt]{4.6pt}{0.5pt}\hspace{-0.5em}\lambda _{N}=\hbar /m_{N}c$ being the
nucleon Compton wavelength ($m_{N\text{ }}$ is the nucleon mass). The
relevant numerical value is $C_{d}^{2}=1.673$. With the aid of (7) and (12),
we can easily express the $S$-matrix parameter $\lambda $ in terms of the
deuteron normalization factor as
\begin{equation}
\lambda =\alpha \frac{C_{d}^{2}+1}{C_{d}^{2}-1}\,.  \tag{13}
\end{equation}
Substituting (13) into (10) and (11), we obtain
\begin{equation}
a=\frac{2}{\alpha }\,\frac{C_{d}^{2}}{C_{d}^{2}+1}\,,  \tag{14}
\end{equation}
\begin{equation}
r_{e}=\frac{1}{\alpha }\,\left( 1-\frac{1}{C_{d}^{2}}\right) \,.  \tag{15}
\end{equation}
Formulas (14) and (15) yield explicit expressions for the low-energy $np$
scattering parameters in terms of the bound-state parameters, the deuteron
binding energy $\varepsilon _{d}$ and the deuteron nuclear vertex constant $%
G_{d}^{2}$. For the scattering length and the effective range, the
substitution of the above experimental values of $\varepsilon _{d}$ and $%
G_{d}^{2}$ into expressions (14) and (15) yields
\begin{equation}
a=5.41\,fm\,,  \tag{16}
\end{equation}
\begin{equation}
r_{e}=1.74\,fm\,.  \tag{17}
\end{equation}
These values are very close to the experimental values \lbrack 11\rbrack
\begin{equation}
a^{expt}=5.42\,fm\,,  \tag{18}
\end{equation}
\begin{equation}
r_{e}^{expt}=1.76\,fm\,.  \tag{19}
\end{equation}
This corresponds to the well-known fact that low-energy neutron-proton
scattering can be very well interpreted by using the effective-range
approximation (9). Thus, low-energy neutron-proton scattering in the triplet
spin state can be accurately described on the basis of data on the deuteron
bound state.

\newpage

\begin{center}
\textbf{3. DERIVATION OF THE VAN OERS-SEAGRAVE FORMULA FROM THE BARGMANN
REPRESENTATION OF THE }$\mathbf{S}$\textbf{\ MATRIX INVOLVING TWO STATES:}
\textbf{NEUTRON-DEUTERON SYSTEM}
\end{center}

\vspace{1pt}

Let us now consider elastic neutron-deuteron scattering in the doublet spin
state at energies below the threshold for deuteron breakup. In this case,
there are two states in the system, the triton ground state and its virtual
state. The existence of the latter has been firmly established \lbrack
12\rbrack . For the case where the system has two states, we take the Jost
function in the form of the rational function
\begin{equation}
f\left( k\right) =\frac{k-i\alpha }{k+i\lambda }\,\frac{k-i\beta }{k+i\mu }%
\,,  \tag{20}
\end{equation}
which has the correct asymptotic behavior given by (2). The parameters $%
\alpha $, $\lambda $, and $\mu $, which appear on the right-hand side of
(20), are positive, while the parameter $\beta $, which corresponds to the
virtual state, is negative. Substituting (20) into (1), we obtain the $S$
matrix for the system having two states in the form
\begin{equation}
S\left( k\right) =\frac{k+i\alpha }{k-i\alpha }\,\frac{k+i\beta }{k-i\beta }%
\,\frac{k+i\lambda }{k-i\lambda }\,\frac{k+i\mu }{k-i\mu }\,.  \tag{21}
\end{equation}
The energies of the bound and the virtual state are given by
\begin{equation}
E_{0}=-\frac{\hbar ^{2}\alpha ^{2}}{2m}\,,  \tag{22}
\end{equation}
\begin{equation}
E_{v}=-\frac{\hbar ^{2}\beta ^{2}}{2m}\,,  \tag{23}
\end{equation}
where the reduced mass $m$ is $(2/3)m_{N}$. The first and the second pole
factor in expression (21) for the $S$ matrix correspond to, respectively,
the bound and the virtual triton state. The third and the fourth factor
involve redundant poles of the $S$ matrix. The nuclear vertex constants $%
G_{T}^{2}$ and $G_{v}^{2}$ for, respectively, the bound and the virtual
state are directly expressed in terms of the residues of the $S$ matrix at
the respective poles $k=i\alpha $ and $k=i\beta $ as
\begin{equation}
G_{T}^{2}=i\pi \rule[5pt]{4.6pt}{0.5pt}\hspace{-0.5em}\lambda ^{2}%
\stackunder{k=i\alpha }{\limfunc{Res}}\,S\left( k\right) \,,  \tag{24}
\end{equation}
\begin{equation}
G_{v}^{2}=i\pi \rule[5pt]{4.6pt}{0.5pt}\hspace{-0.5em}\lambda ^{2}%
\stackunder{k=i\beta }{\limfunc{Res}}\,S\left( k\right) \,,  \tag{25}
\end{equation}
where $\rule[5pt]{4.6pt}{0.5pt}\hspace{-0.5em}\lambda =\frac{3}{2}%
\rule[5pt]{4.6pt}{0.5pt}\hspace{-0.5em}\lambda _{N}$\thinspace . Calculating
the residues and using (24), (25), and (21), we find that the vertex
constants are expressed in terms of the $S$-matrix parameters as
\begin{equation}
G_{T}^{2}=-2\pi \rule[5pt]{4.6pt}{0.5pt}\hspace{-0.5em}\lambda ^{2}\alpha \,%
\frac{\alpha +\beta }{\alpha -\beta }\,\frac{\alpha +\lambda }{\alpha
-\lambda }\,\frac{\alpha +\mu }{\alpha -\mu }\,,  \tag{26}
\end{equation}
\begin{equation}
G_{v}^{2}=2\pi \rule[5pt]{4.6pt}{0.5pt}\hspace{-0.5em}\lambda ^{2}\beta \,%
\frac{\alpha +\beta }{\alpha -\beta }\,\frac{\beta +\lambda }{\beta -\lambda
}\,\frac{\beta +\mu }{\beta -\mu }\,.  \tag{27}
\end{equation}

\vspace{1pt}

For a further analysis, it is worthwhile to introduce the following
combinations of the $S$-matrix parameters:
\begin{equation}
p\equiv \alpha +\beta \,,  \tag{28}
\end{equation}
\begin{equation}
q\equiv \alpha \beta \,,  \tag{29}
\end{equation}
\begin{equation}
u\equiv \lambda +\mu \,,  \tag{30}
\end{equation}
\begin{equation}
v\equiv \lambda \mu \,.  \tag{31}
\end{equation}
Multiplying the corresponding factors in the numerator and the denominator
of (21), we can represent the $S$ matrix in the form
\begin{equation}
S\left( k\right) =\frac{k^{4}-\left( q+v+pu\right) k^{2}+qv+ik\left[ \left(
p+u\right) k^{2}-pv-qu\right] }{k^{4}-\left( q+v+pu\right) k^{2}+qv-ik\left[
\left( p+u\right) k^{2}-pv-qu\right] }\,.  \tag{32}
\end{equation}
Comparing (32) with the representation in (8), we recast the expression for $%
k\cot \delta $ into the form
\begin{equation}
k\cot \delta =\frac{qv-\left( q+v+pu\right) k^{2}+k^{4}}{-pv-qu+\left(
p+u\right) k^{2}}\,.  \tag{33}
\end{equation}
Upon dividing the polynomial in the numerator by the polynomial in the
denominator, we arrive at
\begin{equation}
k\cot \delta =-A+Bk^{2}-\frac{C}{1+Dk^{2}}\,.  \tag{34}
\end{equation}
This expression is nothing but the well-known empirical van Oers-Seagrave
formula \lbrack 4\rbrack , which describes well low-energy neutron-deuteron
scattering in the doublet spin state. The parameters in expansion (34) were
obtained in \lbrack 4\rbrack\ by fitting low-energy experimental data.
Comparing (34) and (33), we express the van Oers-Seagrave parameters in
terms of the $S$-matrix parameters (28)--(31) as
\begin{equation}
A=\frac{pq+uv+pu\left( p+u\right) }{\left( p+u\right) ^{2}}\,,  \tag{35}
\end{equation}
\begin{equation}
B=\frac{1}{p+u}\,,  \tag{36}
\end{equation}
\begin{equation}
C=\frac{qv}{pv+qu}-\frac{pq+uv+pu\left( p+u\right) }{\left( p+u\right) ^{2}}%
\,,  \tag{37}
\end{equation}
\begin{equation}
D=-\frac{p+u}{pv+qu}\,.  \tag{38}
\end{equation}

\vspace{1pt}

Thus, the van Oers-Seagrave formula (34) immediately follows from the
Bargmann representation (21) of the $S$ matrix for the system having two
states. The same two states in the system are responsible for the pole in
the expression for the function $k\cot \delta $. We note that, in the $S$
matrix, it is necessary to take into account, along with two physical poles
corresponding to the bound and the virtual state, two redundant poles, which
ensure the correct asymptotic behavior of the Jost function. For the case of
potentials leading to a finite number of poles in the $S$ matrix, it follows
from the Levinson theorem that the number of redundant poles is determined
by the total number of bound, virtual, and quasistationary states. If we use
an interaction leading to two states in the system, a ground and a virtual
one, the $S$ matrix will therefore have two redundant poles.

\vspace{1pt}

Since the neutron-proton system considered above has a single state, a bound
state in the triplet channel or a virtual one in the singlet channel, the $S$
matrix for a system having one state and the corresponding effective-range
approximation for the function $k\cot \delta $ provide a good approximation
for the $np$ interaction. The limiting transition from the van Oers-Seagrave
formula to the effective-range approximation can easily be obtained by
recasting formula (33) into the form
\begin{equation}
k\cot \delta =\frac{-1/a+c_{2}k^{2}+c_{4}k^{4}}{1+D\,k^{2}}\,.  \tag{39}
\end{equation}
The two-state $S$ matrix (21) reduces to the one-state $S$ matrix (4) if the
second state goes to infinity; that is, $\beta \longrightarrow \infty $ and $%
\mu \longrightarrow \infty $. It can easily be seen that, in this case, the
coefficients $c_{4}$ and $D$ in (39) vanish, so that expression (39) reduces
to the effective-range approximation (9). Thus, we can see that the van
Oers-Seagrave formula is a direct generalization of the effective-range
approximation to the case of a system having two states.

\vspace{1pt}

The nuclear vertex constants $G_{T}^{2}$ and $G_{v}^{2}$ for, respectively,
the bound and the virtual triton state are expressed in terms of the
corresponding dimensionless asymptotic constants $C_{T}^{2}$ and $C_{v}^{2}$
as
\begin{equation}
G_{T}^{2}=3\pi \rule[5pt]{4.6pt}{0.5pt}\hspace{-0.5em}\lambda ^{2}\alpha
C_{T}^{2}\,,  \tag{40}
\end{equation}
\begin{equation}
G_{v}^{2}=3\pi \rule[5pt]{4.6pt}{0.5pt}\hspace{-0.5em}\lambda ^{2}\beta
C_{v}^{2}\,.  \tag{41}
\end{equation}
Using (26), (27), (40), and (41) and taking into account (30) and (31), we
obtain
\begin{equation}
u=-\left( \alpha +\beta \right) \frac{4\left( \alpha +\beta \right)
^{2}+6\left( \alpha ^{2}-\beta ^{2}\right) \left( C_{T}^{2}-C_{v}^{2}\right)
-9\left( \alpha -\beta \right) ^{2}C_{T}^{2}C_{v}^{2}}{4\left( \alpha +\beta
\right) ^{2}-6\left( \alpha +\beta \right) ^{2}\left(
C_{T}^{2}+C_{v}^{2}\right) +9\left( \alpha -\beta \right)
^{2}C_{T}^{2}C_{v}^{2}}\,,  \tag{42}
\end{equation}
\begin{equation}
v=\alpha \beta \frac{4\left( \alpha +\beta \right) ^{2}+6\left( \alpha
+\beta \right) ^{2}\left( C_{T}^{2}+C_{v}^{2}\right) +9\left( \alpha -\beta
\right) ^{2}C_{T}^{2}C_{v}^{2}}{4\left( \alpha +\beta \right) ^{2}-6\left(
\alpha +\beta \right) ^{2}\left( C_{T}^{2}+C_{v}^{2}\right) +9\left( \alpha
-\beta \right) ^{2}C_{T}^{2}C_{v}^{2}}\,.  \tag{43}
\end{equation}
Formulas (35)--(38), together with (28), (29), (42), and (43), provide
explicit expressions for the van Oers-Seagrave parameters in terms of the
parameters of the bound and the virtual triton state (their energies and
nuclear vertex constants). Since these general expressions are very
cumbersome, we consider here only the expressions for the scattering length
and for the pole of the function $k\cot \delta $.

\vspace{1pt}

\begin{center}
\textbf{4. EXPRESSIONS FOR THE }$\mathbf{nd}$\textbf{\ SCATTERING LENGTH AND
FOR THE POLE OF THE FUNCTION }$\mathbf{k}\cot \mathbf{\delta }$\textbf{\ IN
TERMS OF THE PARAMETERS OF THE BOUND AND THE VIRTUAL TRITON STATE}
\end{center}

\vspace{1pt}

As can be seen from (34), the doublet $nd$ scattering length $a\equiv
^{2}a_{nd}$ is expressed in terms of the van Oers-Seagrave parameters as
\begin{equation}
a=\frac{1}{A+C}\,,  \tag{44}
\end{equation}
or in terms of the $S$-matrix parameters as
\begin{equation}
a=\frac{p}{q}+\frac{u}{v}=\frac{1}{\alpha }+\frac{1}{\beta }+\frac{1}{%
\lambda }+\frac{1}{\mu }\,.  \tag{45}
\end{equation}
Substituting (28), (29), (42), and (43) into (45), we obtain
\begin{equation}
a=6\left( \frac{1}{\alpha }+\frac{1}{\beta }\right) \frac{2\left( \alpha
+\beta \right) \left( \beta C_{T}^{2}+\alpha C_{v}^{2}\right) +3\left(
\alpha -\beta \right) ^{2}C_{T}^{2}C_{v}^{2}}{2\left( \alpha +\beta \right)
^{2}\lbrack 2+3\left( C_{T}^{2}+C_{v}^{2}\right) \rbrack +9\left( \alpha
-\beta \right) ^{2}C_{T}^{2}C_{v}^{2}}\,.  \tag{46}
\end{equation}
This formula expresses the doublet $nd$ scattering length in terms of the
parameters of the bound and the virtual triton state. From (38), we
similarly find that the pole $k_{0}^{2}\equiv 1/D$ of the function $k\cot
\delta $ is given by
\begin{equation}
k_{0}^{2}=\alpha \beta \frac{2\left( \alpha +\beta \right) \left( \beta
C_{T}^{2}+\alpha C_{v}^{2}\right) +3\left( \alpha -\beta \right)
^{2}C_{T}^{2}C_{v}^{2}}{2\left( \alpha +\beta \right) \left( \alpha
C_{T}^{2}+\beta C_{v}^{2}\right) -3\left( \alpha -\beta \right)
^{2}C_{T}^{2}C_{v}^{2}}\,.  \tag{47}
\end{equation}

\vspace{1pt}

We note that the effective range for $nd$ scattering can be expressed in
terms of the van Oers-Seagrave parameters as
\begin{equation}
r_{e}=2\left( B+C\,D\right) \,.  \tag{48}
\end{equation}
Along with the effective range, which is anomalously large, $r_{e}\sim
500\,fm$, for doublet $nd$ scattering, the so-called amplitude slope
parameter \lbrack 6\rbrack , given by the dimensionless quantity $\frac{1}{2}%
\alpha _{d}^{3}a^{2}r_{e}$ with wave number $\alpha _{d}=0.2316\,fm^{-1}$
corresponding to the deuteron, is often used in the literature.

\vspace{1pt}

\begin{center}
\textbf{5. NUMERICAL CALCULATION OF THE LOW-ENERGY }$\mathbf{nd}$

\textbf{SCATTERING PARAMETERS AND THE CONSTANT }$C_{v}^{2}$
\end{center}

\vspace{1pt}

With the parameters specifying the bound and the virtual triton state, we
have calculated numerically the low-energy $nd$ scattering parameters $a$, $%
\varepsilon _{0}=-\frac{\hbar ^{2}k_{0}^{2}}{2m}$, and $\frac{1}{2}\alpha
_{d}^{3}a^{2}r_{e}$ by formulas (46)--(48). The calculation has revealed
that these parameters of $nd$ scattering are weakly sensitive to variations
in the energies $E_{0}$ and $E_{v}$ of the ground and the virtual state,
respectively, and to the asymptotic normalization factor $C_{T}^{2}$ for the
bound state. At the same time, these parameters greatly depend on the
asymptotic normalization factor $C_{v}^{2}$ for the virtual state. The
results of the calculations for the low-energy $nd$ scattering parameters
are quoted in the table for various values of the constant $C_{v}^{2}$. In
the calculations, the parameters of the triton were set to the values of $%
E_{T}=8.48\,MeV$ \lbrack 13\rbrack\ ($E_{T}=\left| E_{0}\right| +\varepsilon
_{d}$), $B_{v}=\left| E_{v}\right| =0.482\,MeV$ \lbrack 12\rbrack , and\ $%
C_{T}^{2}=3.5$ \lbrack 1, 2, 6\rbrack , which follow from a direct analysis
of experimental data.

\vspace{1pt}

\vspace{1pt}From the results quoted in the table, it can be seen that the
parameters of $nd$ scattering are highly sensitive to variations in the
constant $C_{v}^{2}$. The experimental value of $C_{v}^{2}=0.0504$ \lbrack
12\rbrack\ adopted at present leads to the $nd$ scattering length $%
a=1.03\,fm $, which differs considerably from its experimental value now
established to a high precision. The experimental value of the $nd$
scattering length \lbrack 14\rbrack ,
\begin{equation}
a^{expt}=0.65\,fm\,,  \tag{49}
\end{equation}
can be fitted with the asymptotic-constant value
\begin{equation}
C_{v}^{2}=0.0592\,,  \tag{50}
\end{equation}
which agrees well with the value of $C_{v}^{2}=0.06$ calculated in the
two-body model of $nd$ interaction simulated by the Hulth\'{e}n potential
\lbrack 2\rbrack . Thus, the numerical value in (50) for the asymptotic
normalization factor $C_{v}^{2}$ for the virtual state must be treated as
the result of the theoretical calculation on the basis of our model. The
high sensitivity of the scattering parameters to this quantity and a
deviation of the theoretical results from its available experimental value
indicate that a refinement of the experimental value of $C_{v}^{2}$ is
necessary. The above highlights the crucial importance of the factor $%
C_{v}^{2}$ for exploring the properties of $nd$ scattering in the doublet
spin state. The experimental value in (49) for the $nd$ scattering length
and, accordingly, the value in (50) for $C_{v}^{2}$ lead\ to the following
values for the pole of the function $k\cot \delta $ and the amplitude slope
parameter:
\begin{equation}
\varepsilon _{0}=-0.1313\,MeV\,,  \tag{51}
\end{equation}
\begin{equation}
\frac{1}{2}\alpha _{d}^{3}a^{2}r_{e}=1.52\,.  \tag{52}
\end{equation}
The value calculated for the pole position $\varepsilon _{0}$ and presented
in (51) agrees well with $\varepsilon _{0}=-0.15\,MeV$ quoted in \lbrack
15\rbrack\ as an experimental value. The slope parameter (52) complies
fairly well with the value of $\frac{1}{2}\alpha _{d}^{3}a^{2}r_{e}=1.35$
\lbrack 6\rbrack\ calculated theoretically from the Faddeev equations with
separable nucleon-nucleon potentials.

\vspace{1pt}

For the van Oers-Seagrave parameters, a numerical calculation with $%
C_{v}^{2} $ from (50) yields
\begin{equation}
A=0.3198\,fm^{-1},  \tag{53}
\end{equation}
\begin{equation}
B=0.7698\,fm\,,  \tag{54}
\end{equation}
\begin{equation}
C=1.2190\,fm^{-1},  \tag{55}
\end{equation}
\begin{equation}
D=236.927\,fm^{2}.  \tag{56}
\end{equation}
The function $k\cot \delta $ calculated in the van Oers-Seagrave
approximation with the parameter values (53)--(56) is shown in the figure
versus the energy $k^{2}$. It can be seen that the theoretical curve
faithfully reproduces experimental data from \lbrack 4\rbrack . In summary,
we have obtained a complete description of low-energy doublet $nd$
scattering in terms of the parameters of the bound and the virtual triton
state and demonstrated that this description is consistent with experimental
data.

\bigskip

\begin{center}
\textbf{6. CONCLUSION}
\end{center}

\vspace{1pt}

It has been shown that the van Oers-Seagrave formula for doublet $nd$
scattering --- it was originally deduced from a purely empirical
consideration --- immediately follows from the Bargmann $S$-matrix
representation corresponding to the presence of two triton states, a bound
and a virtual one, in the system. That the pole term proves to be necessary
in the function $k\cot \delta $ is also due to the presence of two states in
the system. Our analysis has given simple expressions relating the van
Oers-Seagrave parameters to the $S$-matrix parameters. It has been shown
that the van Oers-Seagrave formula is a direct generalization of the
effective-range approximation to the case where the system being considered
has two states; when one of these states goes to infinity, the former
reduces to the latter.

\vspace{1pt}

For the doublet $nd$ scattering length and the pole of the function $k\cot
\delta $, we have obtained simple explicit expressions in terms of the
energies of the bound and the virtual triton state and the nuclear vertex
constants for these states. The resulting formulas make it possible to
perform numerical calculations and to analyze the parameters of low-energy $%
nd$ scattering versus the parameters of the triton. In particular, the
calculations have revealed that the parameters of $nd$ scattering are highly
sensitive to variations in the asymptotic normalization factor $C_{v}^{2}$
for the virtual triton state. The experimental value of the $nd$ scattering
length is fitted at the value of $C_{v}^{2}=0.0592$, which differs from the
experimental value presently accepted for this constant. Therefore, the
constant $C_{v}^{2}$ is of paramount importance for studying the properties
of the $nd$ system and needs further experimental refinement.

\vspace{1pt}\vspace{1pt}

\bigskip

\begin{center}
REFERENCES
\end{center}

\begin{enumerate}
\item  N. M. Petrov, Yad. Fiz. \textbf{48}, 50 (1988) \lbrack Sov. J. Nucl.
Phys. \textbf{48}, 31(1988)\rbrack .

\item  Yu. V. Orlov, N. M. Petrov, and G. N. Teneva, Yad. Fiz. \textbf{55},
38 (1992) \lbrack Sov. J. Nucl. Phys. \textbf{55}, 23 (1992)\rbrack .

\item  L. M. Delves, Phys. Rev. \textbf{118}, 1318 (1960).

\item  W. T. H. van Oers and J. D. Seagrave, Phys. Lett. \textbf{24B}, 562
(1967).

\item  J. S. Whiting and M. G. Fuda, Phys. Rev. \textbf{C 14}, 18 (1976).

\item  I. V. Simenog, A. I. Sitnichenko, and D. V. Shapoval, Yad. Fiz.
\textbf{45}, 60 (1987) \lbrack Sov. J. Nucl. Phys. \textbf{45}, 37
(1987)\rbrack .

\item  V. Bargmann, Rev. Mod. Phys. \textbf{21}, 488 (1949).

\item  R. G. Newton, \textit{Scattering Theory of Waves and Particles}, 2nd
ed. (Springer-Verlag, New York, 1982).

\item  L. D. Blokhintsev, I. Borbeli, and E. I. Dolinski\u{\i}, Fiz. Elem.
Chastits At. Yadra \textbf{8}, 1189 (1977) \lbrack Sov. J. Part. Nucl.
\textbf{8}, 485 (1977)\rbrack .

\item  A. G. Sitenko, \textit{Scattering Theory} (Springer-Verlag, Berlin,
1991).

\item  O. Dumbrajs, R. Koch, H. Pilkuhn, \textit{et al.}, Nucl. Phys.
\textbf{B 216}, 277 (1983).

\item  B. A. Girard and M. G. Fuda, Phys. Rev. \textbf{C 19}, 579 (1979).

\item  D. R. Tilley, H. R. Weller, and H. H. Hasan, Nucl. Phys. \textbf{A 474%
}, 1 (1987).

\item  W. Dilg, L. Koester, and W. Nistler, Phys. Lett. \textbf{36B}, 208
(1971).

\item  Yu. V. Orlov, Yu. P. Orevkov, and L. I. Nikitina, Izv. Akad. Nauk,
Ser. Fiz. \textbf{60} (11), 152 (1996).
\end{enumerate}

\newpage

\bigskip \bigskip

\begin{center}
Low-energy $nd$ scattering parameters versus $C_{v}^{2}$\bigskip

\begin{tabular}{|c|c|c|c|}
\hline
$C_{v}^{2}$ & $a,\,fm$ & $\varepsilon _{0},\,MeV$ & $\frac{1}{2}\alpha
_{d}^{3}a^{2}r_{e}$ \\ \hline
$0.01$ & $3.12$ & $-0.43$ & $0.51$ \\ \hline
$0.02$ & $2.54$ & $-0.38$ & $0.80$ \\ \hline
$0.03$ & $2.01$ & $-0.32$ & $1.04$ \\ \hline
$0.04$ & $1.51$ & $-0.26$ & $1.24$ \\ \hline
$0.05$ & $1.05$ & $-0.20$ & $1.40$ \\ \hline
$0.06$ & $0.62$ & $-0.13$ & $1.53$ \\ \hline
$0.07$ & $0.21$ & $-0.05$ & $1.63$ \\ \hline
$0.08$ & $-0.17$ & $0.04$ & $1.72$ \\ \hline
$0.09$ & $-0.52$ & $0.13$ & $1.79$ \\ \hline
\end{tabular}
\end{center}

\newpage 

\vspace*{2cm}

\begin{center}
\unitlength=0.24pt
\begin{picture}(1225,1599)
\put(0,1599){\special{em:graph 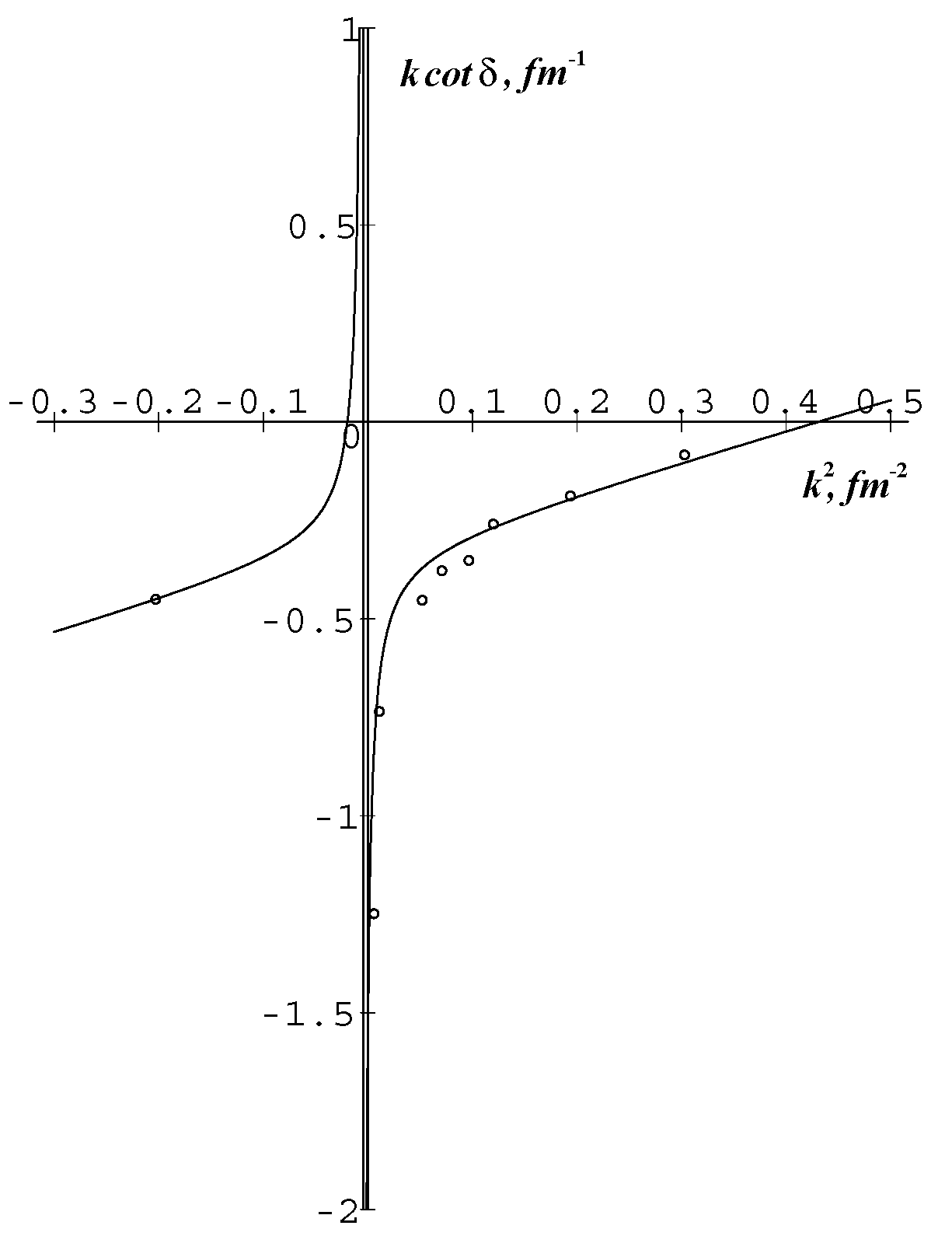}}
\end{picture}
\end{center}

\vspace{12pt}

\noindent Function $k\cot \delta $ calculated for doublet $nd$ scattering in
the van Oers-Seagrave approximation with the parameter values (53)--(56)
versus the energy $k^{2}$. The experimental data were taken from \lbrack
4\rbrack .

\end{document}